\documentclass[letterpaper]{emulateapj}%%%%%% 
\usepackage{amsmath}
\usepackage{graphics}
\usepackage{amsmath,cases}
\usepackage{color}
\newcommand{\msun}{{M}_{\odot}}
\newcommand{\rsun}{{R}_{\odot}}
\newcommand{\zsun}{{Z}_{\odot}}
\newcommand{\gaia}{{\it Gaia}}

\shorttitle{Metallicity dependence of BH-MS}
\shortauthors{KINUGAWA \& YAMAGUCHI.}

\begin{document}

%% LaTeX will automatically break titles if they run longer than
%% one line. However, you may use \\ to force a line break if
%% you desire.

\title{metallicity dependence of black hole main sequence binaries detectable with Gaia}

%% Use \author, \affil, and the \and command to format
%% author and affiliation information.
%% Note that \email has replaced the old \authoremail command
%% from AASTeX v4.0. You can use \email to mark an email address
%% anywhere in the paper, not just in the front matter.
%% As in the title, use \\ to force line breaks.

\author{Tomoya Kinugawa\altaffilmark{1}, and Masaki S. Yamaguchi\altaffilmark{2}}
%\affil{
%Institute for Cosmic Ray Research, The University of Tokyo, 5-1-5 Kashiwa-no-ha, Kashiwa City, Chiba, 277-8582, Japan
%}
%\author{C. D. Biemesderfer\altaffilmark{4,5}}
%\affil{National Optical Astronomy Observatories, Tucson, AZ 85719}
%\email{aastex-help@aas.org}
%
%\and
%
%\author{R. J. Hanisch\altaffilmark{5}}
%\affil{Space Telescope Science Institute, Baltimore, MD 21218}

\altaffiltext{1}{Department of Astronomy, Faculty of Science, The University of Tokyo, 7-3-1, Hongo, Bunkyo-ku, Tokyo, 113-8654 Japan}

\altaffiltext{2}{Department of Physics, Faculty of Science and Engineering, Konan University, 8-9-1 Okamoto, Kobe, Hyogo,658-0072, Japan}

\begin{abstract}

LIGO has detected gravitational waves from massive binary black hole mergers. In order to explain the origin of such massive stellar-mass black holes, extreme metal poor stars including first stars have been invoked.
However, black holes do not carry information of the metallicity.
In order to check the metallicity dependence of the black hole formation, we focus on galactic black hole-main sequence binaries (BH-MSs).
Using a binary population synthesis method, we find that $\gaia$ can detect $\sim200-400$ BH-MSs whose metallicity is $\zsun$ and $\sim70-400$ BH-MSs whose metallicity is $0.1\zsun$. 
With the spectroscopic observation on 4-m class telescopes, we can check the metallicity of BH-MSs.
The metallicity dependence of the black hole formation might be checked by the  astrometry and spectroscopic observations.

\end{abstract}
\section{introduction}

There are some observations of low metallicity main sequence stars at the present day \cite{??}.
Black hole-main sequence binaries (BH-MSs) which are low metal stars also be able to survive at the present day. 
The astrometry observatory $\gaia$ is a powerful tool for search BH-MSs.
Some researchers expected the $\gaia$ can detect $10^2-10^5$ BH-MSs
\citep{Mashian2017,Breivik2017,Yamaguchi2018,Yalinewich2018}.
Such BH-MSs might be mixed with the low metallicity BH-MSs.
%We estimate the detection rate of GAIA for low metal BH-MSs.
With the spectroscopic observation on 4-m class telescopes, we can check the metallicity of BH-MSs. The main sequence star has the information of the metallicity which then allows us to infer the metallicity of the progenitor of the black hole primary.
This is very useful to check the metallicity dependence of black hole formation.

LIGO has detected gravitational waves from binary black holes mergers \citep{Abbot2016a,Abbot2016b,Abbot2016c}.
Some black holes of the gravitational wave sources are $\sim 30~\msun$.
On the other hand, the masses of black hole candidates of X-ray binaries 
are typically $\sim 10~\msun$ \citep{Ozel2010}.
Thus, the origin of massive black holes might be different from general black holes.
In order to explain the origin of such massive stellar black holes, extreme metal poor stars including first stars have been invoked \citep{Kinugawa2014,Kinugawa2016,Belczynski2016,Hartwig2016,Inayoshi2017,Miyamoto2017}.
However, since black holes do not carry information of the metallicity, it might be difficult to determine the population of black hole progenitors until we detect binary black hole mergers at high redshift \citep{Nakamura2016}.　In order to check the metallicity dependence of the black hole formation, the comparison of the theoretical expectation and the results of$\gaia$ and the spectroscopic observation will be important.

{In this paper, we calculate the binary evolution using the population synthesis method to estimate the BH-MS detection rate.
Especially, we calculate two metallicity cases such as $\zsun$, and $0.1\zsun$ and consider the metallicity dependence of the BH-MSs.
We also check the difference of the properties  between the galactic BH-MSs and the observable BH-MSs.
}
\section{method}
%In order to study the BH-MS progenitor evolution, we have to calculate binary evolutions.
%In the case of the binary evolutions, binary interactions make a stellar evolution to change from single stellar evolution.
%Furthermore, how binary interactions take effect depends on the initial binary parameters.
%Thus, w
We use the Monte Carlo method called as the population synthesis to calculate BH-MS rates.
Our binary population synthesis code is based on the BSE code \citep{Hurley2002}.
We rewrite the wind mass loss rate \citep{Kinugawa2017} and the common envelope parts \citep{Kinugawa2014} of the BSE code.

In the part of the wind mass loss, the Wolf Rayet (WR) stellar wind mass loss and the mass loss rate of the luminous blue variable (LBV) stars are updated as
\begin{equation}
\dot{M}_{\rm WR}=10^{-13}\left(\frac{L}{L_{\odot}}\right)^{1.5}\left(\frac{Z}{Z_{\odot}}\right) ^{0.86}~\msun\rm~yr^{-1},
\end{equation}and
\begin{equation}
\dot{M}_{\rm LBV}=1.5\times 10^{-4}~\msun~\rm yr^{-1},
\end{equation}
where $L$, and $Z$ are the luminosity, and the metallicity, respectively \citep{Belczynski2010}.

{When the stellar radius become large and the stellar surface is captured by the gravitational force of the companion, a mass transfer occurs.}
If the separation rapidly shrinks or the stellar radius rapidly expands during the mass transfer, the mass transfer becomes dynamical unstable.
In this case, the companion plunges into the envelope of the donor giant star. 
This phase is called as the common envelope phase.
In this phase, the orbit shrinks and the donor envelope evaporates. 
After this phase, there is a merged star or a binary which consists of the companion and the core of the donor giant star.
%The criterion whatever the binary merges or not is whatever the separation after the common envelope is less than the sum of stellar radii, and whatever the donor star is a giant or not.
For the common envelope phase:
\begin{equation}
\alpha\left(\frac{GM_{\rm{c,1}}M_2}{2a_{\rm{f}}}-\frac{GM_1M_2}{2a_{\rm{i}}}\right)=\frac{GM_{\rm{1}}M_{\rm{env,1}}}{\lambda R_1},
\label{eq:ce1}
\end{equation} 
where $M_1,~M_{\rm c,1},~M_{\rm env,1},~M_2,~a_{\rm i}$, and $a_{\rm f}$ are the mass of the donor giant star, the mass of the donor giant's core, the mass of the donor giant's envelope, the mass of the companion star, the initial separation, and the separation after the common envelope phase, respectively \citep{Webbink1984}.
We use $\alpha\lambda=1$.
{After the CE phase, if the separation $a_{\rm f}$ is less than the sum of the giant's core radius and the companion stellar radius or the donor giant is a Hertzsprung gap phase, the binary will merge \citep{Belczynski2007}}.

At the end of star evolution, we need to calculate the black hole mass.
{In order to calculate the BH mass, we use the equation (1) of \cite{Belczynski2002}.}

To calculate a binary evolution, we need initial binary parameters such as primary muss $M_1$, mass ratio $q=M_2/M_1$, separation $a$, and eccentricity $e$.
We use initial distributions as $f(M_1)\propto M_1^{-2.35} ~(5~\msun<M_1<100~\msun)$, $f(q)={\rm const}.~(0.1~\msun/M_1<q<1)$, $f(a)\propto1/a~(a_{\rm min}<a<10^6~\rsun)$, and $f(e)\propto e~(0<e<1)$, where $a_{\rm min}$ is a minimum separation at either star is just filling its Roche lobe.

We calculate $10^5$ binaries for two metallicity models such as $Z=\zsun$ and $Z=0.1\zsun$.
We choose BH-MSs whose periods $P$ are 50 ${\rm days}<P<5~{\rm yrs}$, because%{in this paper we adopt the astrometric satellite \gaia\, whose cadence for each object and nominal mission period are roughly 50 days and 5 years, respectively. 
the cadence of $\gaia$.
%\textcolor{red}{Here, the condition that orbital period is larger than 50 days is more stringent than the condition that the binary orbit is detected with the \gaia\ astrometry for most part of parameter space we focus on in this paper, so that we neglect the latter condition.}
{We consider two star formation model to check the BH-MSs. First, we use the star formation rate as constant ($SFR\sim2.5~\msun\rm /yr$) for 10 Gyrs \citep{Misiriotis2006}, and assume the fraction of solar metallicity stars and subsolar metallicity stars as $\zsun:0.1\zsun=1:1$ \citep{Panter2008,Belczynski2012}.
We call this model as Mix model.
Second, in order to loughly reflect the chemical evolution of galaxy \citep{Prantzos2008}, we use two star formation rates $SFR_{\zsun}=2.5~\msun\rm /yr ~(t=3.6-13.6~Gyr)$, $SFR_{0.1\zsun}=25~\msun\rm/yr ~(t=2.6-3.6 Gyr)$ for $Z=\zsun$ and $Z=0.1\zsun$, respectively. We call this model as ChemiEvo model.
We use the binary fraction $f_{\rm B}=\frac{N_{\rm binary}}{N_{\rm single}+N_{\rm binary}}=0.5$, where $N_{\rm single}$ and $N_{\rm binary}$ are the number of single stars and the number of binaries, respectively.}

\section{result}

 \begin{table*}[!ht]
 	\caption{The numbers of BH-MSs $N_{\rm BHMS}$ whose periods are 50 days $<P<5~\rm yrs$ for $10^5$ binaries, the numbers of such BH-MSs in the entire galaxy $N_{\rm G}$, and the number of BH-MSs detected by \gaia\ $N_{\rm D}$ for each metallicity case.
 	}
 	\label{IDF}
 	\begin{center}
 		\begin{tabular}{ccc}
 			\hline
 			metallicity	& $\zsun$ & 	$0.1\zsun$  \\
 			\hline
 			$N_{\rm BHMS}$ & 1322 (1322) & 2841 (18)\\
 			$N_{\rm G}$ &4057 (8114)& 7898 (7496) \\
 			$N_{\rm D}$ & 225 (449)& 397 (72)\\
 			\hline
 		\end{tabular}\\
 	\end{center}
 \end{table*}
  The number of BH-MSs in the entire galaxy $N_{\rm G}$ for each metallicity is
 \begin{equation}
 N_{\rm G}=\frac{1}{N_{\rm total}}\sum_{i=1}^{N_{\rm BHMS}}\frac{f_{\rm B}}{1+f_{\rm B}}\cdot \frac{SFR}{2}\cdot t_{{\rm life},i}\cdot f_{\rm IMF} ,
 \end{equation}
 {where $N_{\rm BHMS}$, $t_{{\rm life},i}$, and $f_{\rm IMF}$ (=$\int_{5}^{100}M^{-2.35}dM$/$\int_{0.1}^{100}M^{-2.35}dM$)  are the number of BH-MSs  whose periods are 50 days $<P<5~\rm yrs$ for $N_{\rm total}=10^5$ binaries, the life time of the BH-MS, and the IMF normalization factor, respectively.}

\begin{figure*}
    \begin{tabular}{cc}
      \begin{minipage}[t]{0.5\hsize}
        \centering
        \includegraphics[keepaspectratio, scale=0.5]{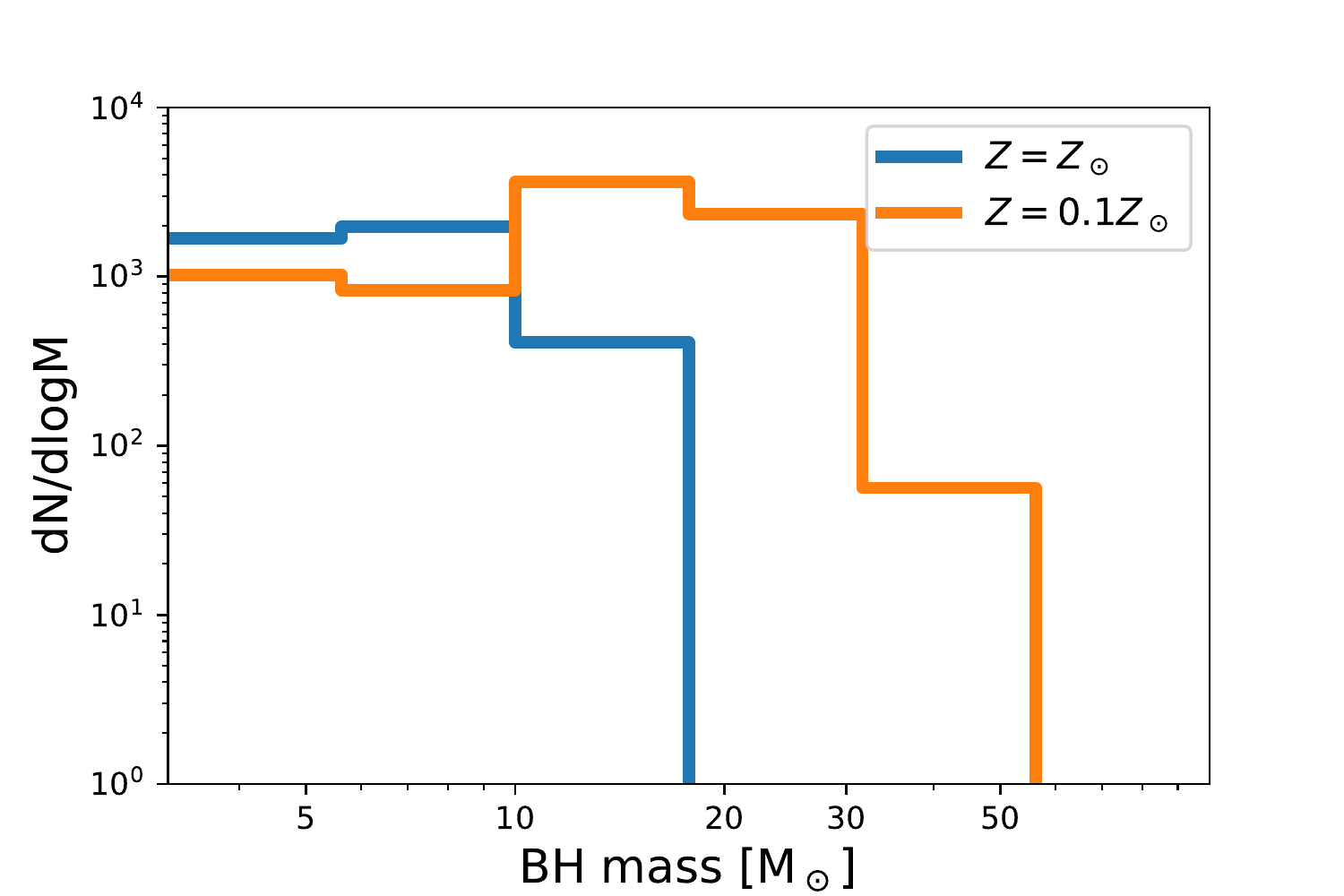}
        \caption{The mass distribution of black holes which are the components of BH-MSs in the entire galaxy for the Mix model.}
        \label{BHmassMix in G}
      \end{minipage} &
      %---- 2番目の図 --------------------------
      \begin{minipage}[t]{0.5\hsize}
        \centering
        \includegraphics[keepaspectratio, scale=0.5]{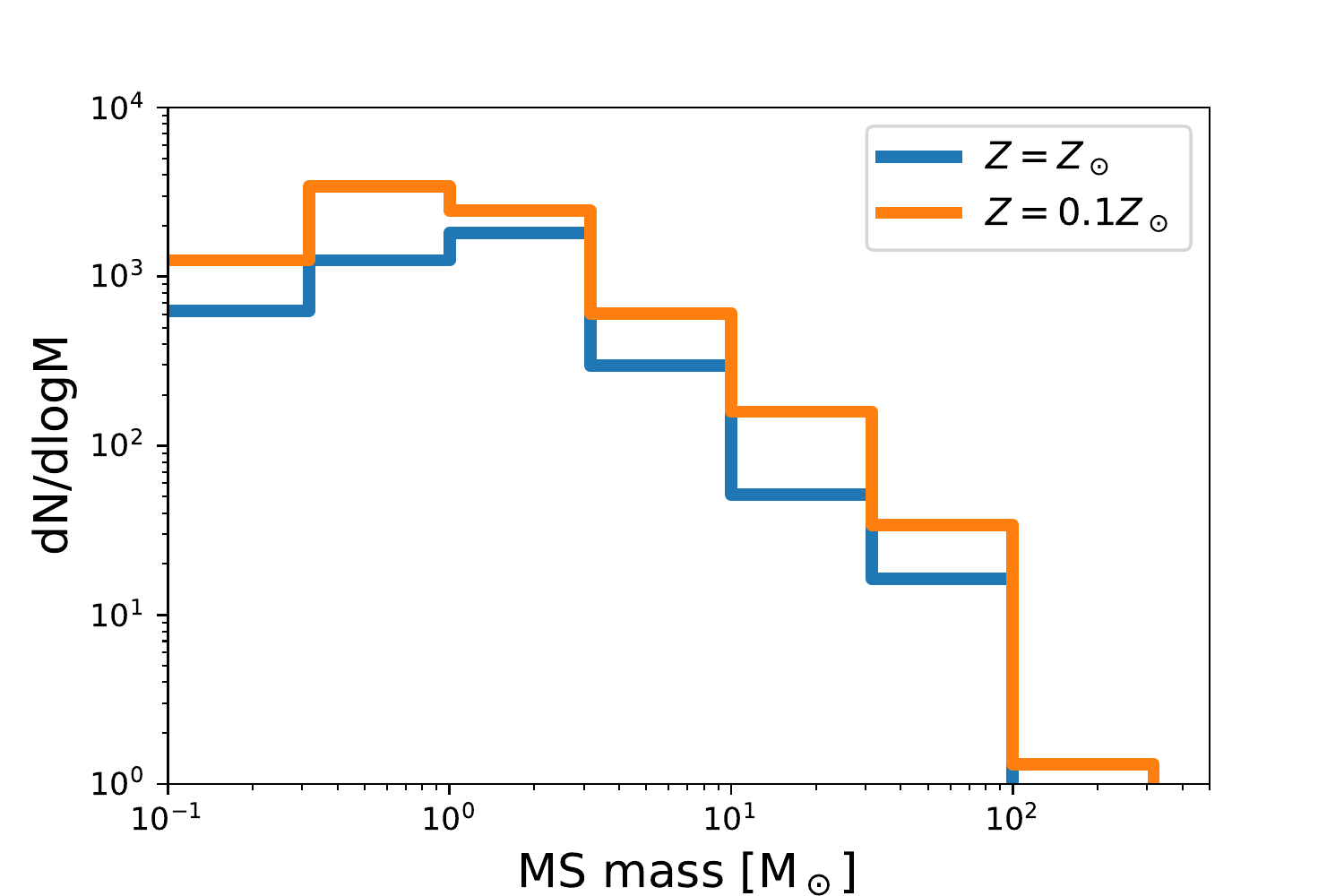}
        \caption{The mass distribution of main sequence stars which are the components of BH-MSs in the entire galaxy for the Mix model.}
        \label{MS massMix in G}
      \end{minipage}
      %---- 図はここまで ----------------------
    \end{tabular}
  \end{figure*}

\begin{figure*}
    \begin{tabular}{cc}
      \begin{minipage}[t]{0.5\hsize}
        \centering
        \includegraphics[keepaspectratio, scale=0.5]{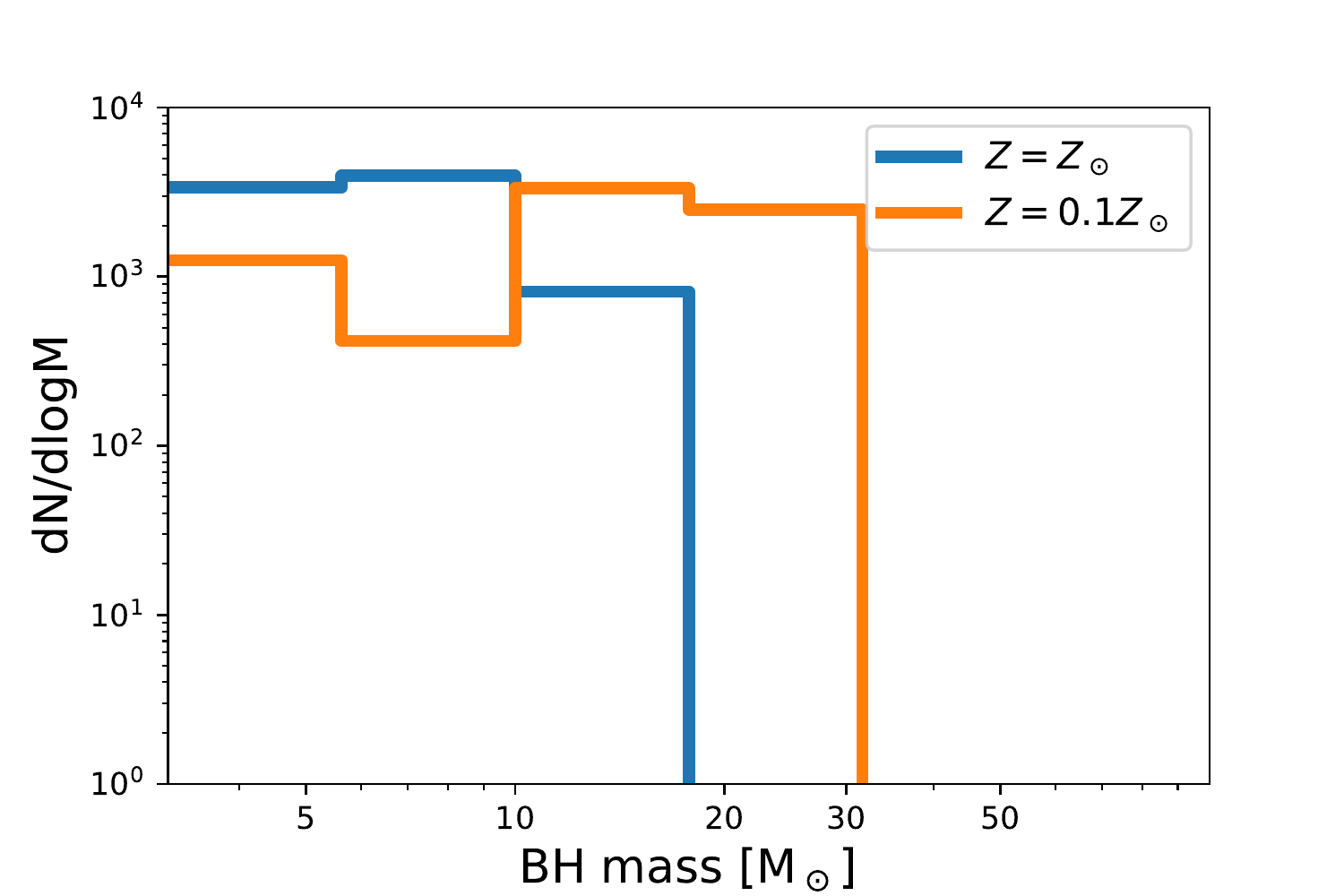}
        \caption{The mass distribution of black holes which are the components of BH-MSs in the entire galaxy for the ChemiEvo model.}
        \label{BHmassChemi in G}
      \end{minipage} &
      %---- 2番目の図 --------------------------
      \begin{minipage}[t]{0.5\hsize}
        \centering
        \includegraphics[keepaspectratio, scale=0.5]{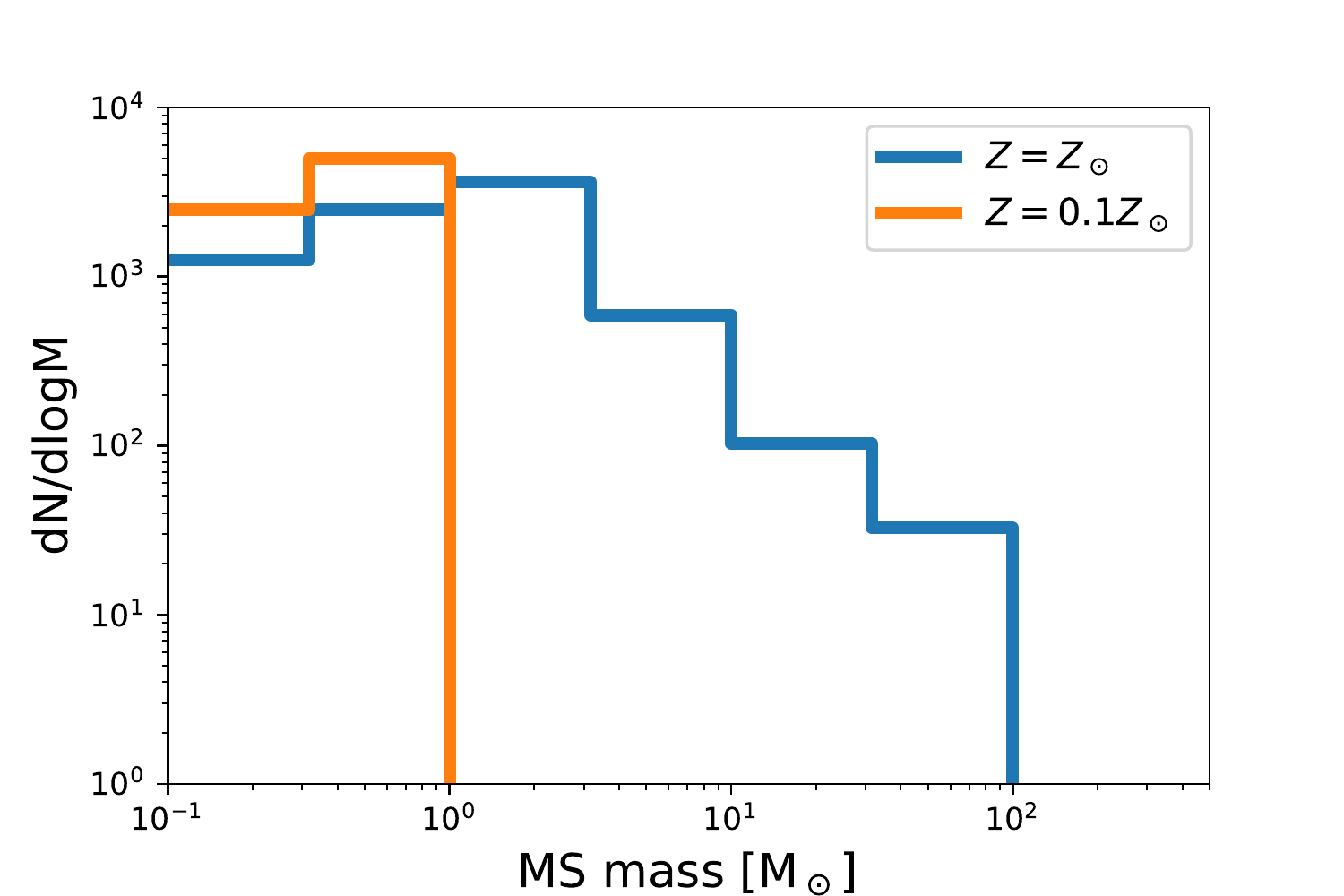}
        \caption{The mass distribution of main sequence stars which are the components of BH-MSs in the entire galaxy for the ChemiEvo model.}
        \label{MS massChemi in G}
      \end{minipage}
      %---- 図はここまで ----------------------
    \end{tabular}
  \end{figure*}

{Figure \ref{BHmassMix in G} and \ref{BHmassChemi in G} show the mass distribution of black holes which are the components of BH-MSs in the entire galaxy for the Mix model and the ChemiEvo model, respectively..
Figure \ref{MS massMix in G} and \ref{MS massChemi in G} show the mass distribution of main sequence stars which are the components of BH-MSs in the entire galaxy for the Mix model and the ChemiEvo model, respectively.
Mass distributions of main-sequence companions do not show dependence on the metallicity in the Mix model.
However, in the ChemiEvo model, there is no massive main-sequence star due to the short lifetime.  
On the other hand, the black hole mass distributions show that the maximum mass for $Z=10\%\zsun$ is clearly more massive then that for  $Z=\zsun$ for each model. }
 
 In order to calculate the number of BH-MSs detected by \gaia, we assume the spatial distribution of BH-MSs in the entire galaxy as  
 \begin{equation}
 \rho_{\rm BHMS}=\rho_0  \exp\left(-\frac{z}{h_z}-\frac{r-r_0}{h_r} \right)
 \end{equation}
where $\rho_0$, $z$, $r$, $r_0~(=8.5~\rm kpc)$, $h_z~(=250~\rm pc)$, and $h_r(=3.5~\rm kpc)$ are the normalization factor of the spatial distribution, the distance perpendicular to the galactic plane, the distance from the galactic center, the distance from the galactic center to the sun, the scale length for the exponential stellar distribution perpendicular to the galactic plane, and the scale length for the exponential stellar distribution parallel to the galactic plane, respectively.
The normalization factor of the spatial distribution is calculated by
\begin{equation}
\rho_0^{-1}={\int_{0}^{\infty}dr\int_{0}^{\infty} \exp\left(-\frac{z}{h_z}-\frac{r-r_0}{h_r} \right)}.
\end{equation}
We use the spherical coordinate centered at the earth, $(D,b,l)$, as
\begin{eqnarray}
r&=&[r_0^2+D^2\cos^2b-2Dr_0\cos b\cos l]^{1/2},\\
z&=&D\sin b,
\end{eqnarray}
where $D$, $b$, and $l$ is the distance from, respectively, the earth, the galactic latitude, and the galactic longitude.
The number of BH-MSs detected by \gaia\ $N_{\rm D}$ is calculated by
\begin{eqnarray}
N_{\rm D}&=&N_{\rm G}  
\times \int_0^{2\pi} dl\int_0^{\pi/2}\cos bdb\int_0^{D_{\rm max}(M)} D^2dD \rho_0,
\end{eqnarray}
where $D_{\rm max}(M) $ is the maximum detectable distance of the BH-MS whose main sequence mass is $M$.

{Here, we derive $D_{\rm max}(M) $ including the signal-to-noise ratio of distance and the precision of astrometric
measurement with \gaia. Generally, the absolute magnitude in the $V$-band $M_V$ can be represented with the apparent magnitude in $V$-band $m_V$ and $D$: $M_V = M_V(m_V,D)$ \citep[e.g., see Eq. (26) in][]{Yamaguchi2018}, where we assume the interstellar extinction $A_V = D/1\rm{kpc}$. In addition, the main sequence mass $M$ can be assumed to be a function of $M_V$: $M(M_V)$  \citep[e.g., Eq. 27 in][]{Yamaguchi2018}. Thus, the main sequence mass is represented as $M(m_V,D)$. 
When we impose the condition that the signal to noise ratio exceeds 10 for the reliable distance measurement, we obtain the condition equation $D/1\rm{kpc} < 10^2/\sigma_\pi(G)$, where $\sigma_\pi$ is the precision of parallax measurement in $G$-band in unit of micro-arcsecond \citep{Gaia2016}. By equating the $G$-band magnitude with $V$-band magnitude, which is justified in \citet{Yamaguchi2018}, we obtain $D_{\rm max} = 10^2\sigma_\pi(m_v)$.
We derive the function $D_{\rm max}(M)$ by solving this equation and $M(m_V,D_{\rm max})$ as simultaneous equations.
}

{Table 1 shows the numbers of BH-MSs $N_{\rm BHMS}$ whose periods are 50 days $<P<5~\rm yrs$ for $10^5$ binaries, the numbers of such BH-MSs in the entire galaxy $N_{\rm G}$, and the number of BH-MSs detected by \gaia\ $N_{\rm D}$ for each metallicity case.
The numbers of binaries with $Z=0.1\zsun$ are about twice larger than the those with $Z=\zsun$ in the Mix model.
There are two reasons.
First, since the mass loss is not so effective in the low metallicity case, they can evolve to more massive compact object.
Second, the low metallicity binaries are more hard to merge within a common envelope phase than the high metallicity binaries, because the Hertzsprung gap phase of the low metallisity stars is shorter than that of the high metallicity \citep{Belczynski2010b}.
On the other hand, the number of binaries ditactable by $\gaia$ with $Z=0.1\zsun$ are about six times less than that of $Z=\zsun$ in the ChemiEvo model.
This reason is that the BH-MSs which consist of massive MS companions are already died due to a short lifetime.}

\begin{figure*}
    \begin{tabular}{cc}
      \begin{minipage}[t]{0.5\hsize}
        \centering
        \includegraphics[keepaspectratio, scale=0.5]{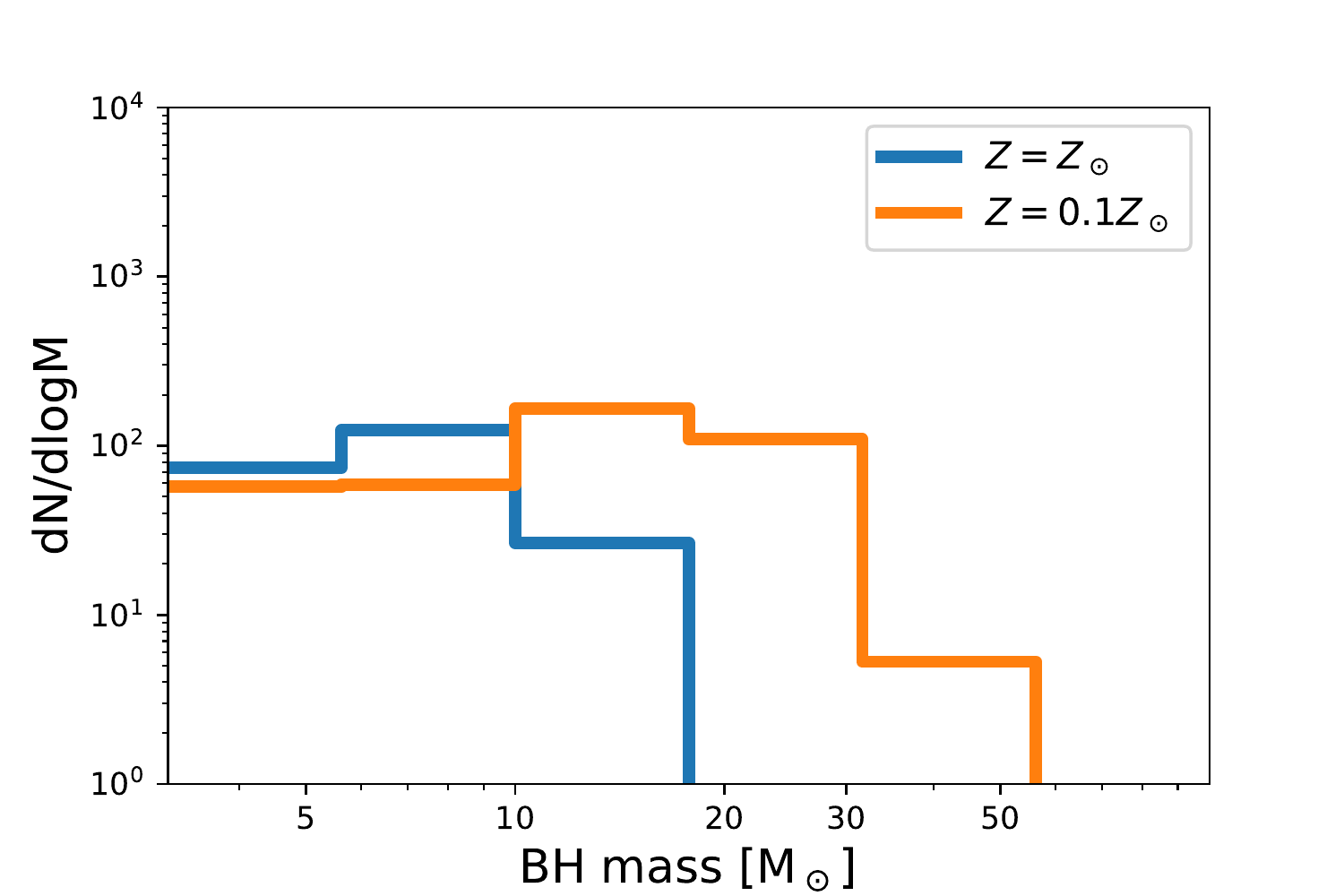}
        \caption{The mass distribution of black holes which are the components of BH-MSs to be detected by \gaia for the Mix model.}
        \label{BHdetectmix}
      \end{minipage} &
      %---- 2番目の図 --------------------------
      \begin{minipage}[t]{0.5\hsize}
        \centering
        \includegraphics[keepaspectratio, scale=0.5]{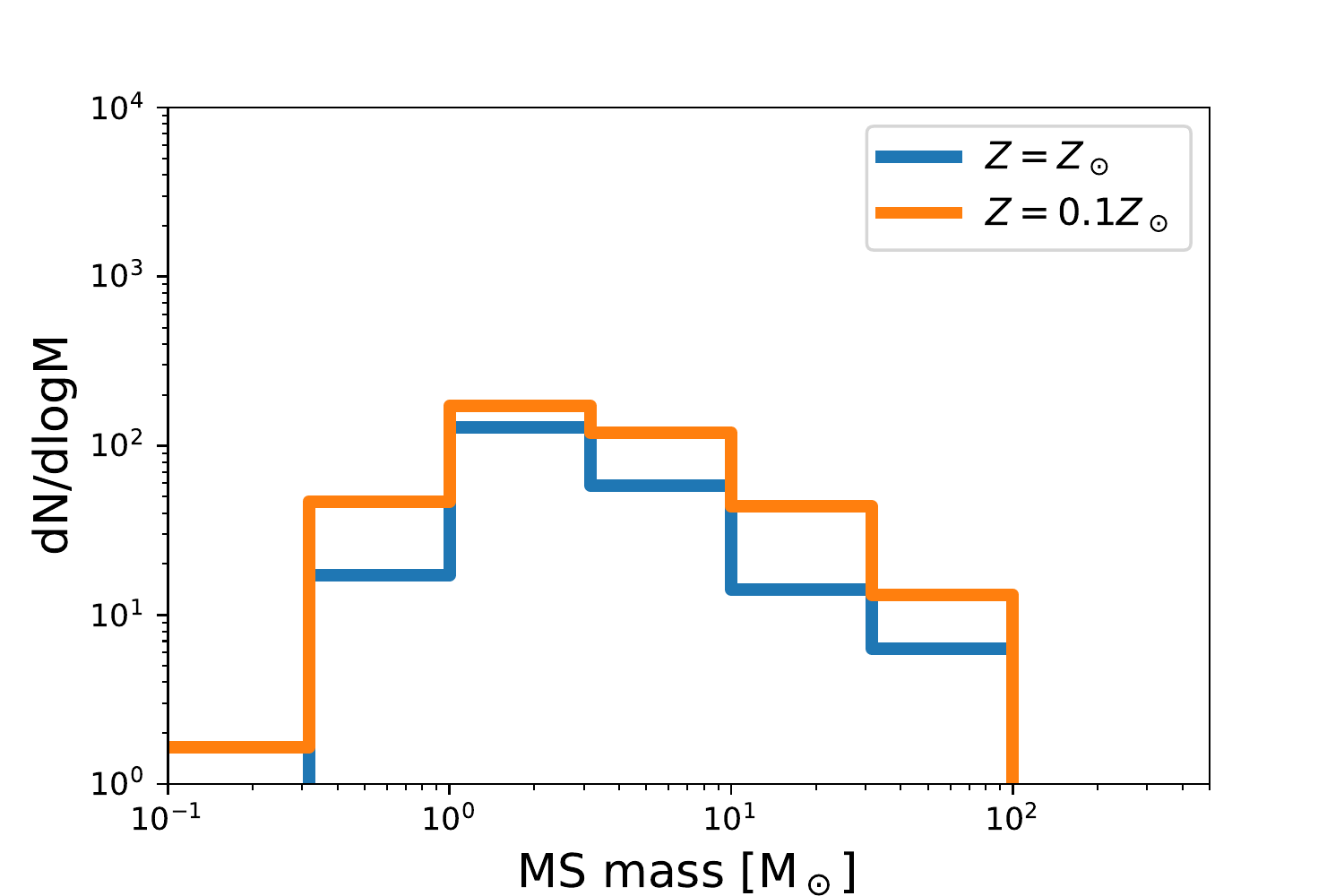}
        \caption{The mass distribution of main sequence stars which are the components of BH-MSs to be detected by \gaia for the Mix model.}
        \label{MSdetectmix}
      \end{minipage}
      %---- 図はここまで ----------------------
    \end{tabular}
  \end{figure*}

\begin{figure*}
    \begin{tabular}{cc}
      \begin{minipage}[t]{0.5\hsize}
        \centering
        \includegraphics[keepaspectratio, scale=0.5]{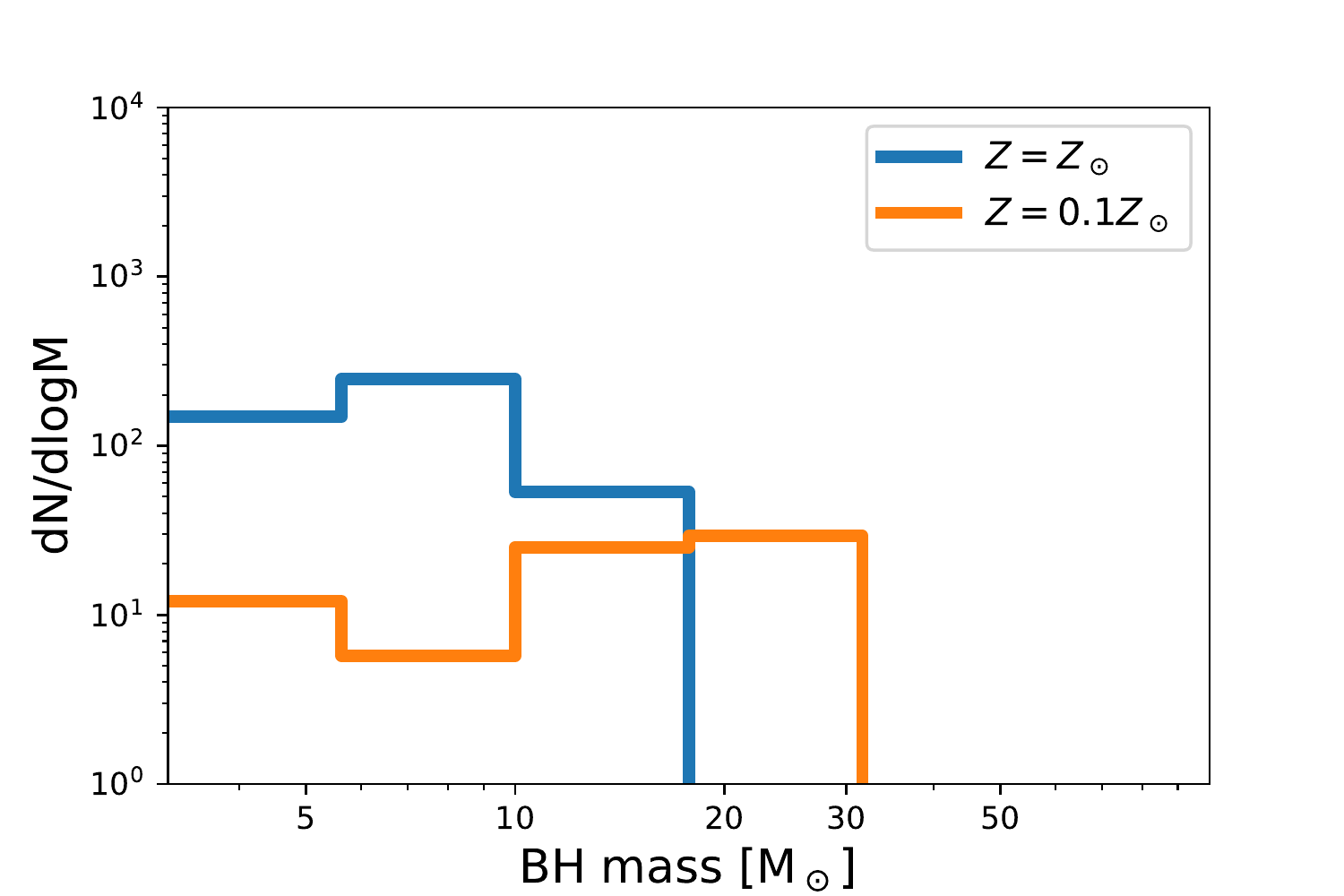}
        \caption{The mass distribution of black holes which are the components of BH-MSs to be detected by \gaia for the ChemiEvo model.}
        \label{BHdetectchemi}
      \end{minipage} &
      %---- 2番目の図 --------------------------
      \begin{minipage}[t]{0.5\hsize}
        \centering
        \includegraphics[keepaspectratio, scale=0.5]{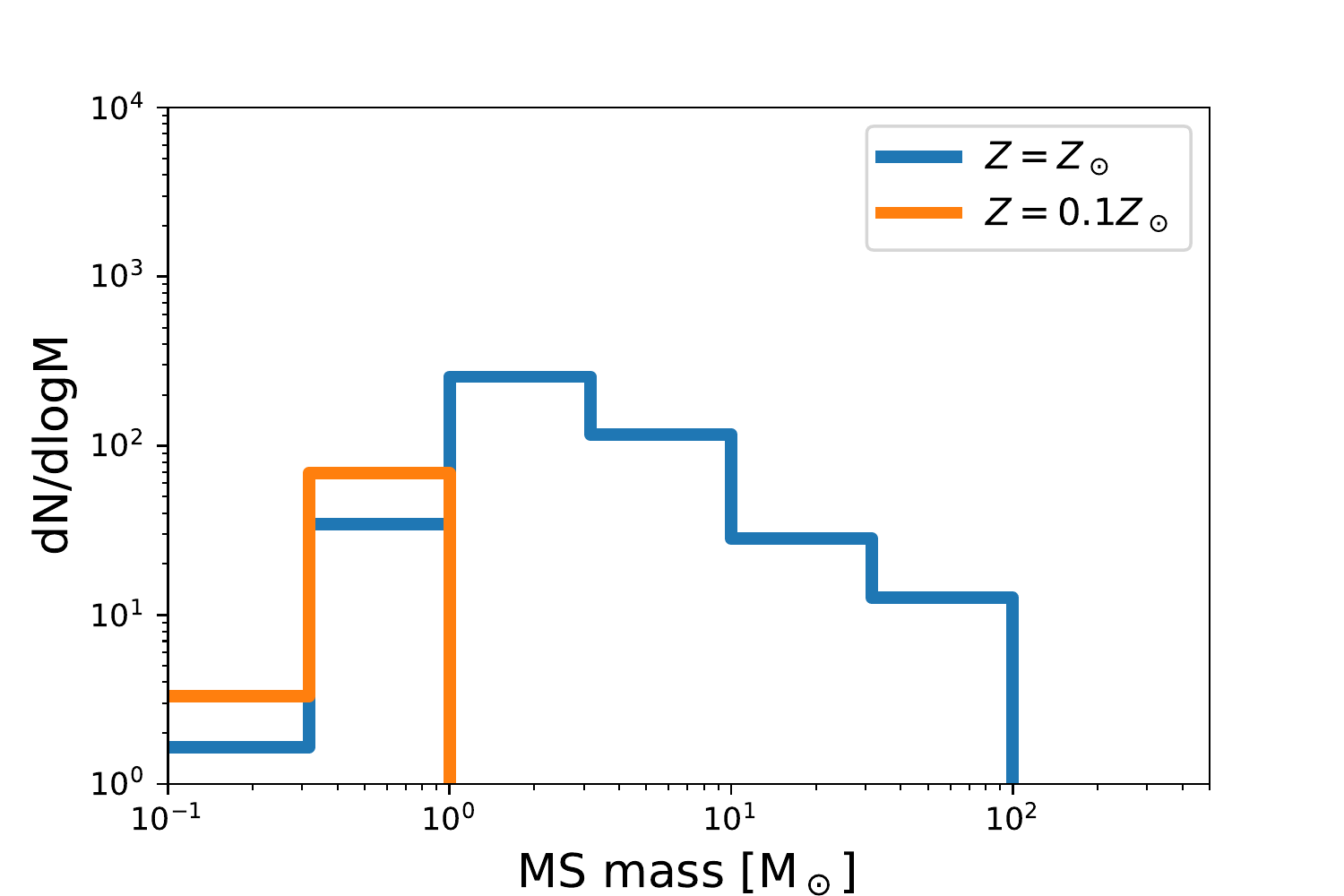}
        \caption{The mass distribution of main sequence stars which are the components of BH-MSs to be detected by \gaia for the ChemiEvo model.}
        \label{MSdetectchemi}
      \end{minipage}
      %---- 図はここまで ----------------------
    \end{tabular}
  \end{figure*}

{Figure \ref{BHdetectmix} and \ref{BHdetectchemi} show mass distributions of black holes in BH-MSs detected by $\gaia$ for the Mix model and the ChemiEvo model.
Figure \ref{MSdetectmix} and \ref{MSdetectchemi} show mass distributions of main sequence stars in BH-MSs detected by $\gaia$ for the Mix model and ChemiEvo model. 
As with Fig. \ref{BHmassMix in G} and Fig. \ref{BHmassChemi in G},  black hole mass distributions detected by $\gaia$ depend on the metallicity. The mass distribution of black holes detected by $\gaia$ is almost the same in shape as that for the entire galaxy. 
On the other hand, in the case of the main-sequence companion, the number of low mass main-sequence companion is clearly smaller than that of main sequence stars of BH-MSs in the entire galaxy. This is because the low mass components are hard to be detected by $\gaia$ due to their faintness.}

\section{Conclusion and Discussion}
We find that $\gaia$ can identify $\sim 600$ galactic BH-MSs.
This number is almost the same as the BH-MS detection number of \cite{Yamaguchi2018}, which is a few times lager than that of \cite{Yalinewich2018}.
%On the other hand, that of \cite{Breivik2017} is 6-30 times larger than our result.
%Furthermore, that of \cite{Mashian2017} is $\sim300$ times larger than our result.}
%?The main difference between our calculation and \cite{Breivik} is the criterion of $\gaia$ detection.?
%In the case of \cite{Mashian2017}, the criterion of $\gaia$ detection and treatment of binary evolution make this difference.}
We consider the BH-MS detection fraction by $\gaia$ for two metallicity cases.
Figure \ref{BHdetectmix}, \ref{BHdetectchemi} show that the BH mass distributions obviously depend on the metallicity. 
If the BH-MS whose BH mass is more massive than $\sim18~\msun$ is detected by $\gaia$, it might be Pop II origin.

{The metallicity of each binary can be measured with a follow-up spectroscopic observation. A main-sequence star in Pop II binaries should show a low metallicity and its typical brightness is expected to be $V \lesssim$20 mag, as the limiting magnitude of Gaia is 20 magnitude in G-band. Thus, this low metallicity can be measured with spectroscopic observations using 4m-class telescopes, such as AAT (Anglo-Australian Telescope) at NSW, Mayall telescope at Kitt Peak, and Kyoto university 3.8m telescope at Okayama.}

If the metallicity of black hole progenitor is revealed by the $\gaia$ astrometry and follow-up observations, we can know the metallicity dependence of the black hole mass distribution. Although the gravitational wave observations have revealed the existence of the massive stellar black hole and will describe the black hole mass distribution, they cannot reveal the metallicity of the progenitor. $\gaia$ can be the powerful tool for the research of the black hole progenitor study. 
\section*{Acknowledgment}
%%%%%%%%%%%%%%%%%%%%%%%%%%%%%%%%%%%%%%%
We thank Shri Kulkarni and Michiko Fujii for useful comments and discussion. This research was supported by JSPS KAKENHI grant numbers JP18J00558(TK), JP17H06363 and JP18K13576(MSY).

%%%%%%%%%%%%%%%%%%%%%%%%%%%%%%%%%%%%%%%%

%%%%%%%%%%%%%%%%%%%%%%%%%%%%%%%%%%%%%%%%
\end{document}